# Metamaterial-Inspired Quad-Band Notch Filter for LTE Band Receivers and WPT Applications

R. Keshavarz[(1)], Y. Miyanaga[(2)], M. Yamamoto[(2)], T. Hikage[(2)], and N. Shariati[(1),(2)]
(1) University of Technology Sydney, Ultimo, NSW 2007, Australia.
(2) Hokkaido University, GI-CoRE, GSB, Kita 14, Nishi9, Kita-Ku, Sapporo, Hokkaido, 0600814 Japan.

## Abstract

A new compact quad-band notch filter (QBNF) based on the extended composite right and left-handed transmission line (E-CRLH TL) has been presented. As known, E-CRLH TL behaves like a quad-band structure. A microstrip TL which is loaded with an open-ended E-CRLH TL is presented as a QBNF. Four unwanted frequencies were used in a dual-band LTE receiver as four notch frequencies which must be eliminated (0.9 GHz, 1.3 GHz, 2.55 GHz, and 3.35 GHz). Also, this QBNF can be applied to simultaneous wireless power and data transfer (SWPDT) system to isolate the wireless power circuit from the data communication circuit. A design technique for the proposed QBNF is presented and its performance is validated using full-wave simulation results and theoretical analysis. The main advantage of this design is an overall rejection greater than 20dB at selected unwanted frequencies. Good agreements between the full-wave simulation and equivalent circuit model results have been achieved which verified the effectiveness of the proposed circuit model. The proposed QBNF is designed on an FR-4 substrate and the dimension of the proposed QBNF is 20×22 mm.

## 1 Introduction

The architectures of wireless receivers can generally be divided into two categories; homodyne and heterodyne receivers. A homodyne receiver, also known as a direct-conversion receiver (DCR), is an RF receiver configuration that down-convert the incoming RF signal using a local oscillator whose frequency is identical to, or very close to the carrier frequency of the input RF signal. On the other hand, in the heterodyne architecture, the signal goes through a receiver chain where its frequency translated to a lower intermediate frequency (IF) [1].

One main drawback in the heterodyne receiver is the problem of image frequency ($f_{IM}$). Image frequency is an undesired input frequency equal to $f_{RF} + 2f_{IF}$ (if $f_{RF} < f_{LO}$) or $f_{RF} - 2f_{IF}$ (if $f_{RF} > f_{LO}$). The image frequency can be received at the same time with the RF signal and down-converted into the IF frequency, thus producing interference in the receiver. Image frequencies can be eliminated by sufficient attenuation in the incoming signal using RF notch filter as an image rejection filter (IRF) of the heterodyne receiver; then, signal processing operations are performed [1].

Beside the image component, in the high-sensitive receiver, if the local oscillator generates second-order harmonic frequency ($2f_{LO}$), then the unwanted frequency $f_{SH} = 2f_{LO} + f_{IF}$ can be down-converted into the desired IF frequency band. $f_{SH}$ is an interference frequency due to the second harmonic generation of the local oscillator.

Consequently, these undesired signals ($f_{IM}$, $f_{SH}$) distort the down-converted desired signal ($f_{IF}$), if they are not filtered away before the down-conversion. As a result, we need a multi-band notch filter to eliminate these unwanted frequencies.

Another application of a multi-band notch filter is using in wireless power transfer (WPT) and energy harvesting (EH) systems [2]. The dual-functional WPT system as simultaneous wireless power and data transfer (SWPDT) is an ideal candidate for several applications such as Internet-of-Things (IoT) that require simultaneous wireless information and power transfer. In this structure, wireless power circuit configured to receive wireless power from the antenna at a first frequency, communication circuit coupled to the antenna and configured to receive a signal from the antenna at a second frequency different from the first frequency [3], and a notch filter between the antenna and an input of the wireless power circuit is needed to isolate the wireless power circuit from the data communication circuit.

In recent years, different techniques have been employed to design notch filters [4], [5] and [6]. Most recently the appearance of composite right and left-handed (CRLH) lines has opened new possibilities in the design of dual-band structures [7]. Moreover, combining the conventional CRLH (C-CRLH) and the dual CRLH (D-CRLH) into an extended CRLH (E-CRLH) prototype paves the way for novel arbitrary quad-band components. The procedure to design a quad-band structure has been analytically explained [8].

In this paper, a new quad-band notch filter (QBNF) based on the E-CRLH TL is presented. The equivalent circuit model of an E-CRLH TL consists of a series-LC resonant circuit and a parallel-LC resonant circuit in both vertical and horizontal branches. A microstrip line is used which is loaded with an open-ended E-CRLH TL which can behave as a QBNF. The proposed configuration can remove interferer frequencies ($f_{IM}$, $f_{SH}$) in two different RF frequency bands ($f_{RF1}$, $f_{RF2}$) and can be used in a dual-band receiver.

This paper is organized as follows: Section II presents a theoretical description and design principle of the proposed QBNF based on the E-CRLH TL. In section III, proposed QBNF is simulated and its results are presented and compared with analytical results. Finally, conclusions are presented in section IV.

## 2 Theoretical Analysis of the Proposed Quad-Band Notch Filter (QBNF)

According to [8], by combining the conventional CRLH and the dual CRLH (D-CRLH), an extended CRLH (E-CRLH) with a quad-band performance can be realized. Figure 1 shows a unit cell E-CRLH TL and its equivalent circuit model. Series impedance and parallel admittance for this transmission line can be derived as follows:

$$Z = j\omega L_R^c/2\left(1 - \frac{\omega_{cR}}{\omega^2}\right) - \frac{j}{\omega C_L^d/2\left(1 - (\omega_{dR}/\omega^2)\right)}$$

$$Y = j\omega C_R^c\left(1 - \frac{\omega_{cL}}{\omega^2}\right) - \frac{j}{\omega L_L^d\left(1 - (\omega_{dL}/\omega^2)\right)} \quad (1)$$

where

$$\omega_{cs} = 1/\sqrt{L_R^c C_L^c}, \quad \omega_{dp} = 1/\sqrt{L_R^d C_L^d}$$
$$\omega_{cp} = 1/\sqrt{L_L^c C_R^c}, \quad \omega_{ds} = 1/\sqrt{L_L^d C_R^d} \quad (2)$$

The Bloch propagation constant is equal to [8]:

$$\cos(\beta p) = 1 + ZY \quad (3)$$

$p$ is the physical length of the unit cell.
Also, the Bloch impedance of E-CRLH TL is given by [8]:

$$Z_B = \sqrt{\frac{Z}{Y}}\sqrt{2 + ZY} \quad (4)$$

and around the $\beta = 0$ frequencies can be approximated by:

$$Z_B = \sqrt{\frac{2Z}{Y}} \quad (5)$$

A quad-band (QB) device is a component accomplishing the same function at four different arbitrary frequencies $\omega_1, \omega_2, \omega_3$ and $\omega_4$. Such a component is therefore constituted of TL sections inducing equivalent phase shifts:

$$\varphi_i = \beta_i p \quad i = 1, 2, 3 \text{ and } 4 \quad (6)$$

It was shown that an E-CRLH TL can exhibit similar properties of a quad-band component and its analytical equations for designing procedures were presented in [8].

A typical heterodyne receiver is shown in figure 2. As can be seen, the notch filter is an essential part of this architecture. As mentioned earlier, without rejection of undesired signals in the operational frequency band, the desired RF and unwanted signals are both mixed down to IF frequency.

In order to achieve high performance of the heterodyne receiver, we use QBNF to cancel four interferers ($f_{IM1}, f_{SH1}, f_{IM2}, f_{SH2}$) frequencies at two RF frequencies ($f_{RF1}, f_{RF2}$) in a dual-band receiver.

The schematic of the proposed QBNF is shown in figure 3.a. In this figure, a conventional microstrip TL has been loaded with an open-ended E-CRLH TL which exhibits a phase difference of $|\pi/2|$ in four notch frequencies. Therefore, the proposed structure exhibits band-stop property in four frequency bands and can be used as a QBNF.

Consequently, in the design procedure, firstly we design an E-CRLH TL based on the derived equations in [8] which exhibit phase difference of $|\pi/2|$ in four unwanted frequencies. Then, by using this E-CRLH TL as an open-ended loaded into a conventional microstrip TL, a QBNF is implemented.

## 3 Simulation Results

The proposed QBNF is designed on the FR-4 substrate with 1.6 mm thickness, loss tangent equal to 0.03 and dielectric constant of 4.7, as shown in figure 3.b. This structure is simulated by ADS (Advanced Design System) software.

We select two LTE frequency sets for RF and LO in an assumed dual-band receiver:

$$\begin{cases} f_{RF1} = 1.4 \; GHz, & f_{LO1} = 1.15 \; GHz, & f_{IF1} = 250 \; MHz \\ f_{RF1} = 1.8 \; GHz, & f_{LO1} = 1.55 \; GHz, & f_{IF1} = 250 \; MHz \end{cases} \quad (7)$$

According to (7), unwanted frequencies for two RF frequencies are:

$$set1: \begin{cases} f_{IM1} = 0.9 \; GHz \\ f_{SH1} = 2.55 \; GHz \end{cases} \quad set2: \begin{cases} f_{IM2} = 1.3 \; GHz \\ f_{SH2} = 3.35 \; GHz \end{cases} \quad (8)$$

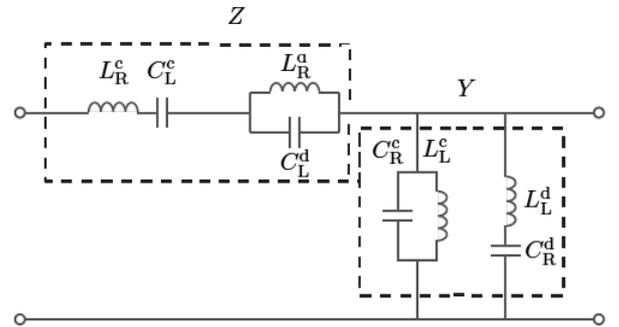

**Figure 1.** Equivalent circuit model of E-CRLH TL.

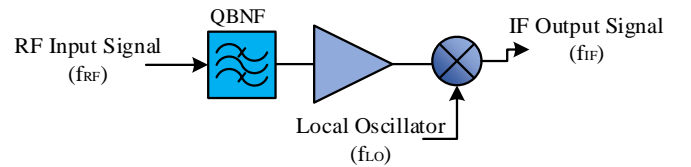

**Figure 2.** Heterodyne receiver architecture.

Unwanted frequencies in (8) can be eliminated in a dual-band receiver (figure 2) by using the proposed QBNF. The values of the equivalent circuit model components (figure 1) have been calculated from [8] and optimized in ADS software and then presented in Table 1.

Figure 4 shows the S-parameters of the proposed QBNF. It is evident that a quad-band notch filtering performance has been achieved around the design frequencies of 0.9GHz, 1.4GHz, 2.55GHz, and 3.6GHz. The rejection of the proposed QBNF in notch frequencies is more than 20dB, which proofs the good performance of the designed IRF at desired frequencies.

Table 1. Lumped elements values of figure 1.

| $C_R^c$ | $L_L^c$ | $L_L^d$ | $C_R^d$ | $L_R^c$ | $C_L^c$ | $C_L^d$ | $L_R^d$ |
|---|---|---|---|---|---|---|---|
| 2.6 $pF$ | 3.7 $nH$ | 3.3 $nH$ | 1.9 $pF$ | 6.4 $nH$ | 1.5 $pF$ | 1.3 $pF$ | 4.8$nH$ |

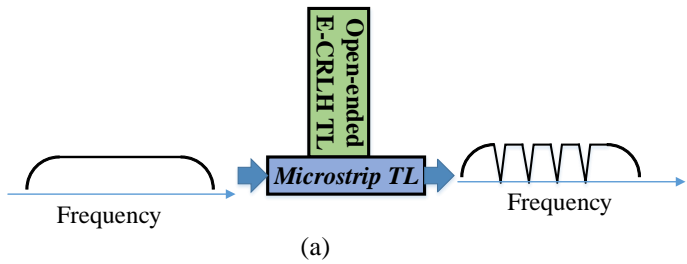

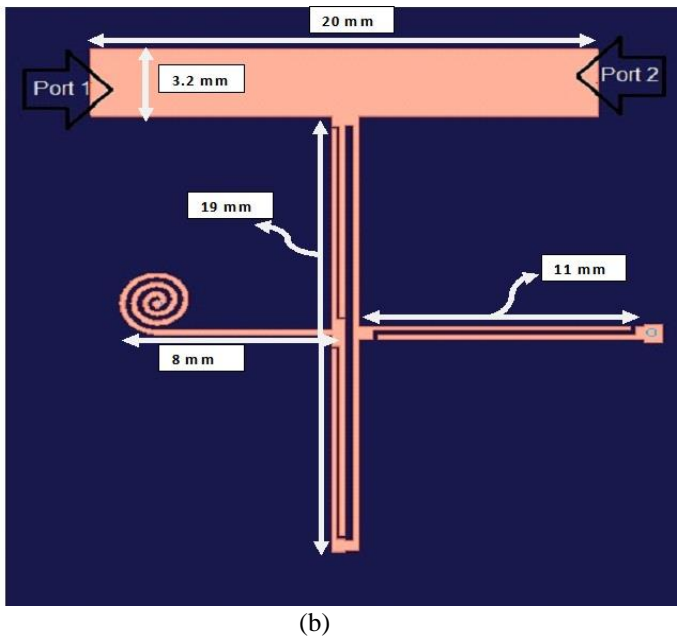

Figure 3. Proposed QBNF on FR4 substrate, a) schematic, b) layout (top view).

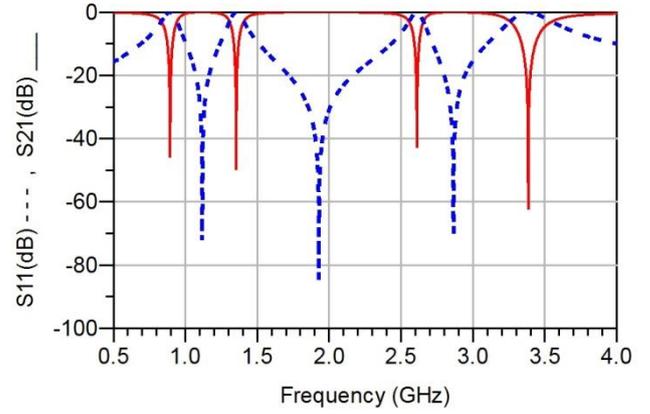

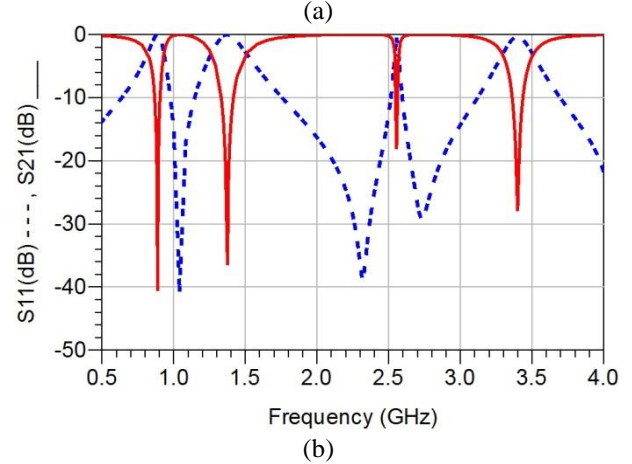

Figure 4. S-parameters of the proposed QBNF, a) theoretical analysis (equivalent circuit model), b) full-wave simulation analysis (layout).

## 4 Conclusion

A new type of quad-band notch filter (QBNF) composed of a conventional microstrip line which is loaded with an open-ended E-CRLH TL has been proposed, investigated theoretically and simulated. The proposed QBNF with a dimension of 20×22 mm exhibits the rejection of 20 dB in four unwanted frequency bands 0.9 GHz, 1.3 GHz, 2.55 GHz, and 3.35 GHz. In order to analyze the introduced QBNF, an equivalent circuit model has been presented and validated by full-wave simulation results. The proposed QBNF can be used as a notch filter in a dual-band LTE receiver configuration or SWPDT system to reject unwanted frequencies.

## 5 Acknowledgments

This study is supported in parts by the Ministry of Education, Science, Sports and Culture, Japan, Grant-in-Aid for Scientific Research, Fund for the Promotion of Joint International Research, Fostering Joint International Research (B) (18KK0277).